\documentclass{article}
\usepackage{spconf}
\usepackage{cite}
\usepackage{adjustbox}
\usepackage[ruled,vlined]{algorithm2e}
\usepackage{amsmath,amssymb,amsfonts}
\usepackage{algorithmic}
\usepackage{graphicx}
\usepackage{textcomp}
\usepackage[colorlinks=false, pdfborder={0 0 0}, breaklinks]{hyperref}
\usepackage{xcolor}
\usepackage{balance}
\usepackage{enumitem}
\setlist{nolistsep,leftmargin=*}
\usepackage{setspace}
\usepackage{wrapfig}
\usepackage{listings}
\usepackage{color}
\usepackage{caption}
\usepackage{subcaption}
\usepackage{todonotes}
\usepackage[longtable]{multirow}

\def\BibTeX{{\rm B\kern-.05em{\sc i\kern-.025em b}\kern-.08em
    T\kern-.1667em\lower.7ex\hbox{E}\kern-.125emX}}

\usepackage{amsmath}

\usepackage{tikz}
\usetikzlibrary{shapes.geometric, arrows,shadows}
\usetikzlibrary{fit}
\usetikzlibrary{calc}
\tikzstyle{process_g} = [rectangle, minimum width=2.3cm, minimum height=0.75cm, text centered, text width=2.3cm, draw=black, fill=gray!40]

\tikzstyle{process} = [rectangle, minimum width=2.3cm, minimum height=0.75cm, text centered, text width=2.3cm, draw=black, fill=gray!10]
\tikzstyle{arrow} = [thick,->,>=stealth]

\usepackage{ctable}
\newcommand{\cf}{\emph{cf.}\xspace}

\newcommand{\ie}{\emph{i.e.}, }
\newcommand{\eg}{\emph{e.g.}, }
\newcommand{\etc}{\emph{etc. }}

\newcommand{\HEVC}{\emph{High Efficiency Video Coding }}
\newcommand{\VVC}{\emph{Versatile Video Coding }}

\begin{document}

\title{All-intra rate control using\\low complexity video features for Versatile Video Coding}

\name{%
\begin{tabular}{@{}c@{}}
Vignesh V Menon$^{1}$ \qquad
Anastasia Henkel$^2$ \qquad
Prajit T Rajendran$^3$ \qquad
Christian R. Helmrich$^2$\\ 
Adam Wieckowski$^{2}$ \qquad 
Benjamin Bross$^{2}$ \qquad 
Christian Timmerer$^{1}$ \qquad 
Detlev Marpe$^{2}$
\end{tabular}\vspace*{-4mm}}
\address{\small $^1$ Christian Doppler Laboratory ATHENA, Alpen-Adria-Universit{\"a}t, Klagenfurt, Austria\\
\small $^2$ Video Communication and Applications Department, Fraunhofer HHI, Berlin, Germany\\
\small $^3$ CEA, List, F-91120 Palaiseau, Université Paris-Saclay, France}

\maketitle

\begin{abstract}
\VVC~(VVC) allows for large compression efficiency gains over its predecessor, \HEVC~(HEVC). The added efficiency comes at the cost of increased runtime complexity, especially for encoding. It is thus highly relevant to explore all available runtime reduction options. This paper proposes a novel first pass for two-pass rate control in  all-intra configuration, using low-complexity video analysis and a Random Forest (RF)-based machine learning model to derive the data required for driving the second pass.
The proposed method is validated using VVenC, an open and optimized VVC encoder. Compared to the default two-pass rate control algorithm in VVenC, the proposed method achieves around 32\% reduction in encoding time for the preset \textit{faster}, while on average only causing $2\%$ BD-rate increase and achieving similar rate control accuracy.
\end{abstract}


\section{Introduction}
\begin{figure}[t]
\centering
\includegraphics[width=.45\textwidth,clip]{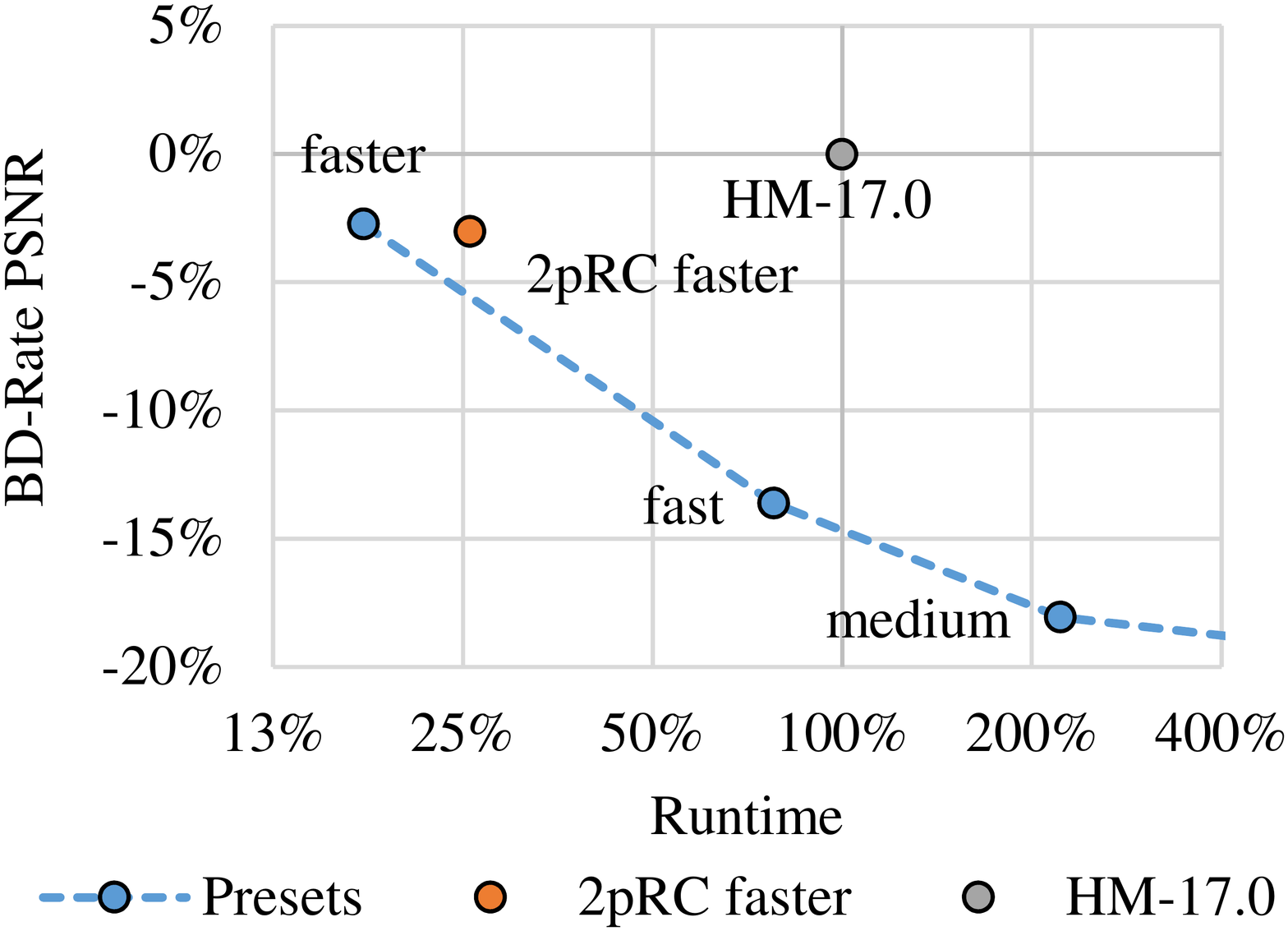}
\caption{VVenC 1.7.0 presets for fixed QP all-intra encoding as well as the working point of two-pass rate control for the
preset \textit{faster}, compared to HM-17.0.}
\vspace{-0.5em}
\label{fig:vvenc_presets}
\end{figure}

Modern video standards come with ever-increasing complexity. The newest, \VVC~(VVC)~\cite{overview_vvc}, was already during its planning intended to be up to ten times more complex to encode than its predecessor, \HEVC~(HEVC)~\cite{overview_hevc}. Practical implementations, like the open source Versatile Video Encoder (VVenC)~\cite{vvenc_ref}, can efficiently deal with scaling the compression efficiency versus the runtime~\cite{preset_ref} as shown in Fig.~\ref{fig:vvenc_presets}, by providing different presets allowing to trade-off runtime against compression efficiency.

\begin{figure*}[t]
\centering
\begin{subfigure}{0.40\textwidth}
    \centering
    \includegraphics[trim={0.3cm 10.5cm 29.50cm 0cm}, clip,width=\textwidth]{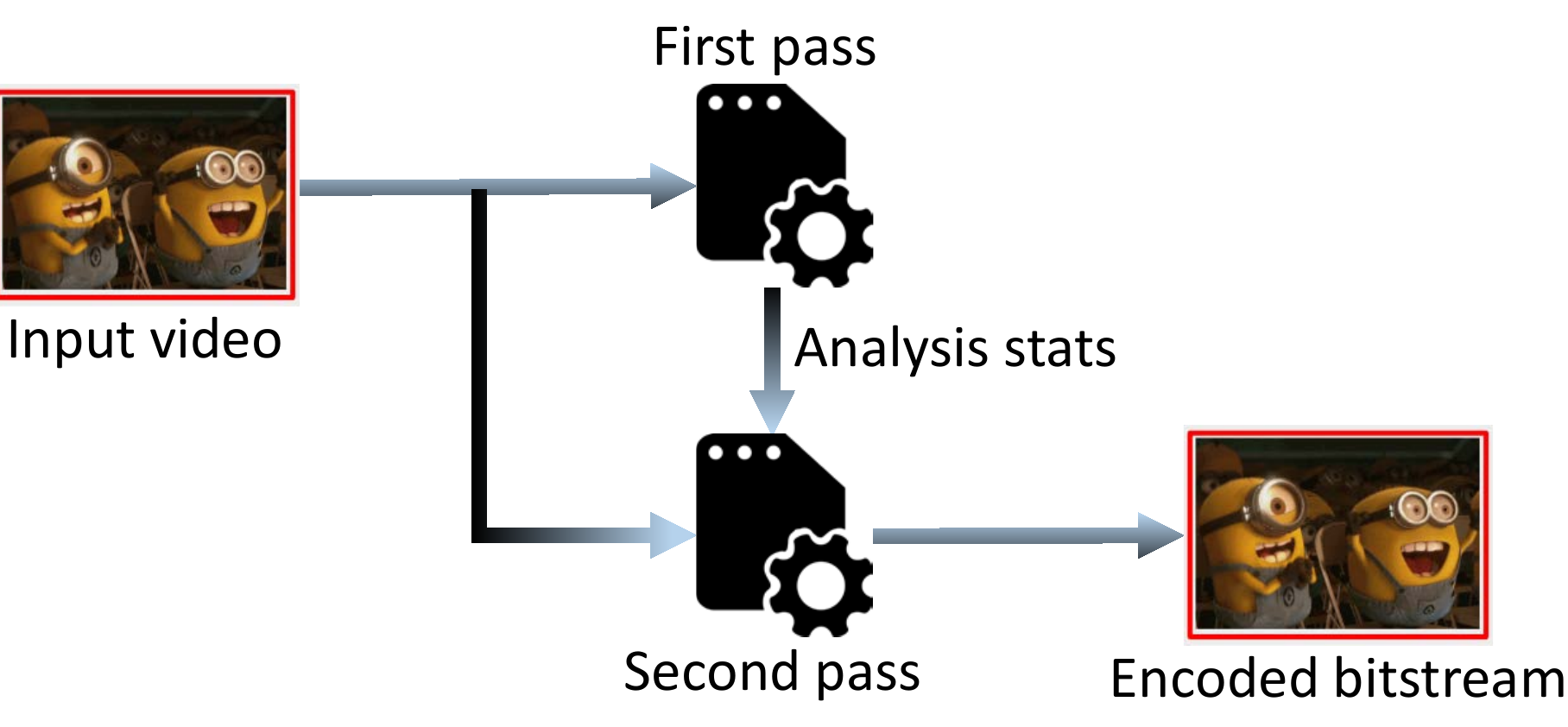}
    \caption{}
    \label{fig:two_pass_arch}
\end{subfigure}
\begin{subfigure}{0.59\textwidth}
    \centering
    \includegraphics[trim={0.0cm 10.2cm 20.90cm 0cm}, clip,width=\textwidth]{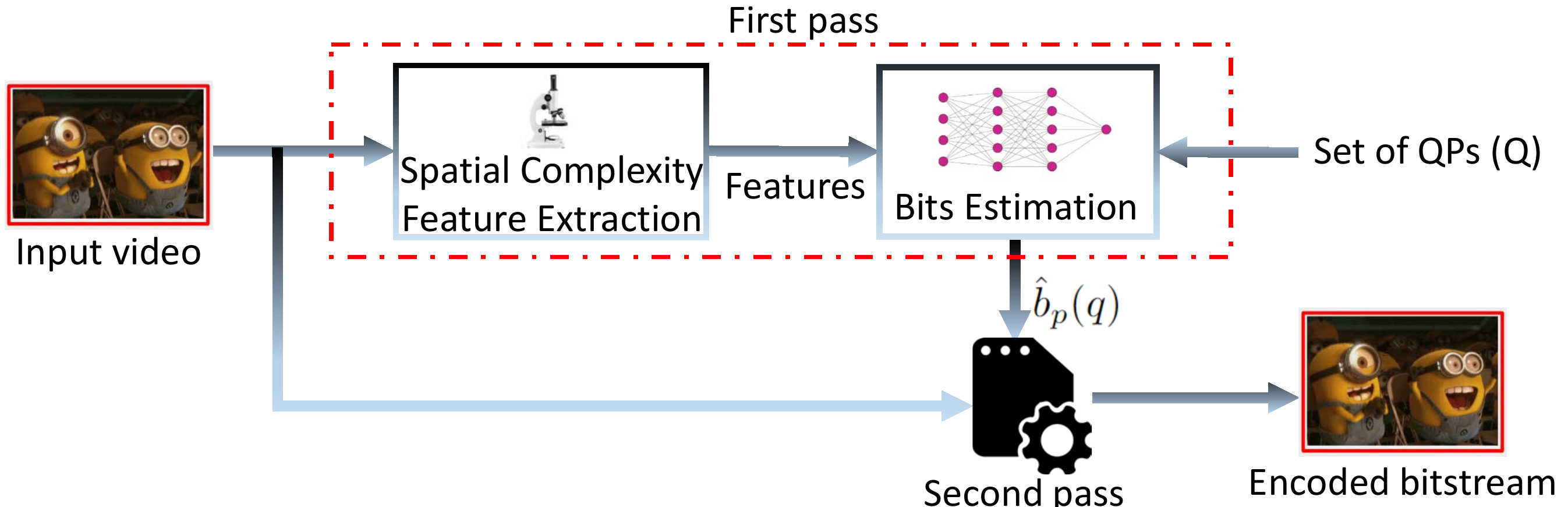}
    \caption{}
    \label{fig:prop_method_arch}    
\end{subfigure}
\caption{(a) State-of-the-art and (b) proposed two-pass rate control encoding architecture.}
\vspace{-0.5em}
\end{figure*}

\textbf{\textit{Motivation:}} With reduced runtime complexity, additional processing steps of constant runtime are increasingly significant. In VVenC, especially the rate control shows this behavior. The encoder processes the input signal twice for the two-pass rate control (2pRC)~\cite{rc_ref,rc_ref2}. The first pass, \ie encoding the video using a fixed reduced toolset, is used to collect basic statistics. Those are then used in the second pass, using the desired working point (or preset), to drive bit allocation between pictures for optimal rate distribution (\cf Fig.~\ref{fig:two_pass_arch}) such that \textit{(i)} the overall quality of the video is constant over time, and  \textit{(ii)} the final encoding has roughly a specific size. While for the high-efficiency presets (\eg \textit{slower} preset) of VVenC, the 2pRC encoding does not take substantially longer than an encoding using a fixed quantization parameter (QP) value resulting in a specific rate, the overhead of the first pass is higher for faster
presets. However, for the preset \textit{faster}, the 2pRC runtime is up to 150\% of a fixed QP encoding resulting in a similar rate. In Fig. \ref{fig:vvenc_presets}, it can be observed that the runtime of 2pRC for preset \textit{faster} lies significantly below the Pareto front at a comparable speed.

All-Intra video coding is a video compression method that encodes each video frame independently, without referring to any previously encoded frames. This is in contrast to the more common inter-frame coding methods, where the encoding of a frame is based on the difference between the current frame and previously encoded frames. All-intra coding is vital because it provides high-quality video, low latency, improved error resilience, random access, and efficient editing. These benefits make all-intra coding useful in live video streaming and professional video production applications. In all-intra encoding, no motion compensation or other inter-frame dependencies are allowed. For example, such a mode can be used as a benchmark for motion-compensated modes~\cite{Ohm2012ComparisonOT}. All-intra coding is also widely used in practice, \eg when encoding single frames (\ie still picture coding) or when instant random access to every frame is required (\eg mezzanine codecs are mostly all-intra). In this paper, the experiments are performed for all-intra coding conditions.

\textbf{\textit{Target:}} This paper aims to minimize the first pass encoding time, \ie time taken to collect statistics used in the second pass while achieving similar rate control accuracy. To this light, this paper explores an alternative method for statistics collection during the first pass to reduce the overall runtime of two-pass rate control encoding, one of the essential encoding modes in state-of-the-art video encoders like x264\footnote{\href{https://www.videolan.org/developers/x264.html}{https://www.videolan.org/developers/x264.html}, last access: Feb 20, 2023.}, x265\footnote{\href{https://www.videolan.org/developers/x265.html}{https://www.videolan.org/developers/x265.html}, last access: Feb 20, 2023.}, and VVenC~\cite{vvenc_ref}. As can be observed in~\cite{preset_ref} for the default use case of motion compensated encoding using VVenC, and in~\cite{vvenc_allintra} for other use cases, a runtime increase of 50\% around the preset \textit{faster} provides a substantial bitrate reduction if following the Pareto front.

\textbf{\textit{Contributions:}} This paper proposes a two-pass rate control method for intra-coding that includes using a light-weight estimator of frame complexity as the first pass as shown in Fig.~\ref{fig:prop_method_arch}. The proposed first pass produces the statistics used in the second encoding pass with lower computational overhead than the default two-pass encoding scheme in the state-of-the-art video encoders. Video Coding Analyzer (VCA)~\cite{vca_ref} is used as the low-complexity estimator. While VCA has proven helpful in estimating encoding complexity~\cite{vca_app, jtps_ref, tqpm_ref, ds_paper}, it does not provide the exact statistics as input to the second pass encoding rate control, \ie the bitrate distribution from the first pass. A Random Forest (RF)-based machine learning model is designed to predict the required number of bits from the VCA complexity estimation to overcome this problem. The proposed two-pass rate control method is validated using VVenC. It achieves around 32\% reduction in encoding time for the preset \textit{faster}, while on average only causing $2\%$ BD-rate increase and achieving similar rate control accuracy compared to the two-pass rate control method in VVenC.

\textbf{\textit{Paper outline:}} Section \ref{sec:proposed_alg} describes the proposed operation of a reduced complexity two-pass rate control for VVenC. In Section \ref{sec:evaluation}, the accuracy of the designed model and the empirical results of the encoding system described in Section \ref{sec:proposed_alg} are evaluated. Section \ref{sec:conclusion_future_dir} concludes the paper.

\section{Two-pass rate control using low-complexity bit estimation}
\label{sec:proposed_alg}    
\subsection{First Pass}
\label{sec:firstpass}
The first pass of the proposed rate control method is divided into three steps: 
\textit{(i)} spatial complexity feature extraction, \textit{(ii)} bits estimation, and \textit{(iii)} application to VVenC.

\textit{(i) Spatial complexity feature extraction:} The commonly used spatial complexity feature is Spatial Information (SI)~\cite{siti_itu_ref}, but its correlation with encoding output features such as bitrate, encoding time, \etc is very low, which is insufficient for encoding parameter prediction~\cite{vca_ref}. In this paper, six DCT-energy-based features are used:
\begin{itemize}[leftmargin=*]
    \item the average luminance texture energy $E_{Y}$;
    \item the average luminance $L_{Y}$;
    \item the average chrominance texture energy $E_{U}$ and $E_{V}$ (for U and V planes); and
    \item the average chrominance $L_{U}$ and $L_{V}$ (for U and V planes).
\end{itemize} These features are extracted using the open source Video Complexity Analyzer (VCA)\footnote{\label{ref_vca}\href{https://vca.itec.aau.at}{https://vca.itec.aau.at}, last access: Feb 15, 2023.}~\cite{vca_ref} and represented as the following vector.

\begin{equation}
x = [E_{Y}, L_{Y}, E_{U}, L_{U}, E_{V}, L_{V}]    
\end{equation}

\textit{(ii) Bits estimation:} For each I-frame, the number of bits is predicted using the spatial features (\ie luminance and chrominance features) of the frame for each quantization parameter $q$. In this paper, a random forest regression model is used.
\begin{equation}
\Tilde{x} = [x | q]^{T}    
\end{equation}
The predicted bits $\hat{b}$ can be presented as $\hat{b}$ = $f(\Tilde{x})$.
The loss function used for training this model is the mean squared error (MSE), which measures the average difference between the predicted and actual (ground truth) number of bits. 


\textit{(iii) Application to VVenC:} For each frame $p$, the above bits
estimator \textit{(ii)} provides a bit-count prediction $\hat{b}_p$
for each possible QP value $q_p$. Then, the first pass
compiles a list of $\hat{b}_p$ values for
each $p$ according to the $q_p$ values assigned
to each frame in the first pass, using $f(\tilde{x})$.

\subsection{Second Pass}

For each frame, the first pass rate-QP estimator of Section~\ref{sec:firstpass} provides a pre-assigned QP value $q_p$ and an associated bit-count prediction $\hat{b}_p$. Using this data pair and the target bit
count $b'_p$ for the second pass, VVenC's $R$-$QP$ model \cite{rc_ref} determines the closest integer QP value $q'_p$ corresponding
to $b'_p$ and performs the second pass encoding using that $q'_p$:
\begin{equation}
\bar{q}_p= q_p-c_\text{low} \cdot \sqrt{\text{max}(1;q_p)} \cdot \text{log}_2 {\biggl(\frac{b'_p}{\hat{b}_p}\biggr)},
\label{eq:eq1}
\end{equation}
where $c_\text{low}$ is a constant for the low-rate end of the $R$-$QP$
function and $\bar{q}_p$ is an initial second pass QP value.
The final $q'_p$ is obtained using a high-rate corrective step as:
\begin{equation}
q'_p= \text{round} {\bigl(\bar{q}_p+c_\text{high} \cdot \text{max}(0; q_\text{start}-\bar{q}_p)\bigr)},
\label{eq:eq12}
\end{equation}
where 0 $<c_\text{high}<$  1 is a video resolution dependent constant
(\eg 0.5 for 2160p and 0.25 for 480p input) and $q_\text{start}$ = 24 was
chosen experimentally. Further details can be found in~\cite{icip23rc}.

Any inaccuracies in the bit consumption of a frame after this final rate-distortion optimal encoding pass (\ie too many or too few bits spent in the final pass) will be accumulated and compensated for in the following frames, as described in~\cite{rc_ref}.
\vspace{-0.3em}
\section{Evaluation}
\label{sec:evaluation}
\subsection{Bits Estimation Model}
\paragraph*{Training and Hyperparameters:}
\label{sec:test_methodology}
The hyperparameters used in the RF model are $random\_state=0$, $min\_samples\_leaf=1$, $min\_samples\_split=2$, and $n\_estimators=100$. For $max\_depth$, four different values, \ie 4, 8, 12, and 16, have been experimented with, considering a trade-off between model size and prediction accuracy. In this paper, prediction accuracy is evaluated using the coefficient of determination ($R^{2}$) score and Mean Absolute Error (MAE) compared to the ground truth values.

To train the bits estimation model, four hundred UHD sequences (80\% of the sequences) from the Video Complexity Dataset~\cite{VCD_ref} are used as the training set, and the remaining (20\%) is used as the validation set. The sequences are encoded at 24 fps using VVenC~v1.7.0\footnote{\label{note:vvenc}\href{https://github.com/fraunhoferhhi/vvenc}{https://github.com/fraunhoferhhi/vvenc}, last access: Feb 20, 2023.}~\cite{vvenc_ref} with the \textit{faster} preset.

\begin{figure}[t]
\centering
    \centering
    \includegraphics[width=0.45\textwidth]{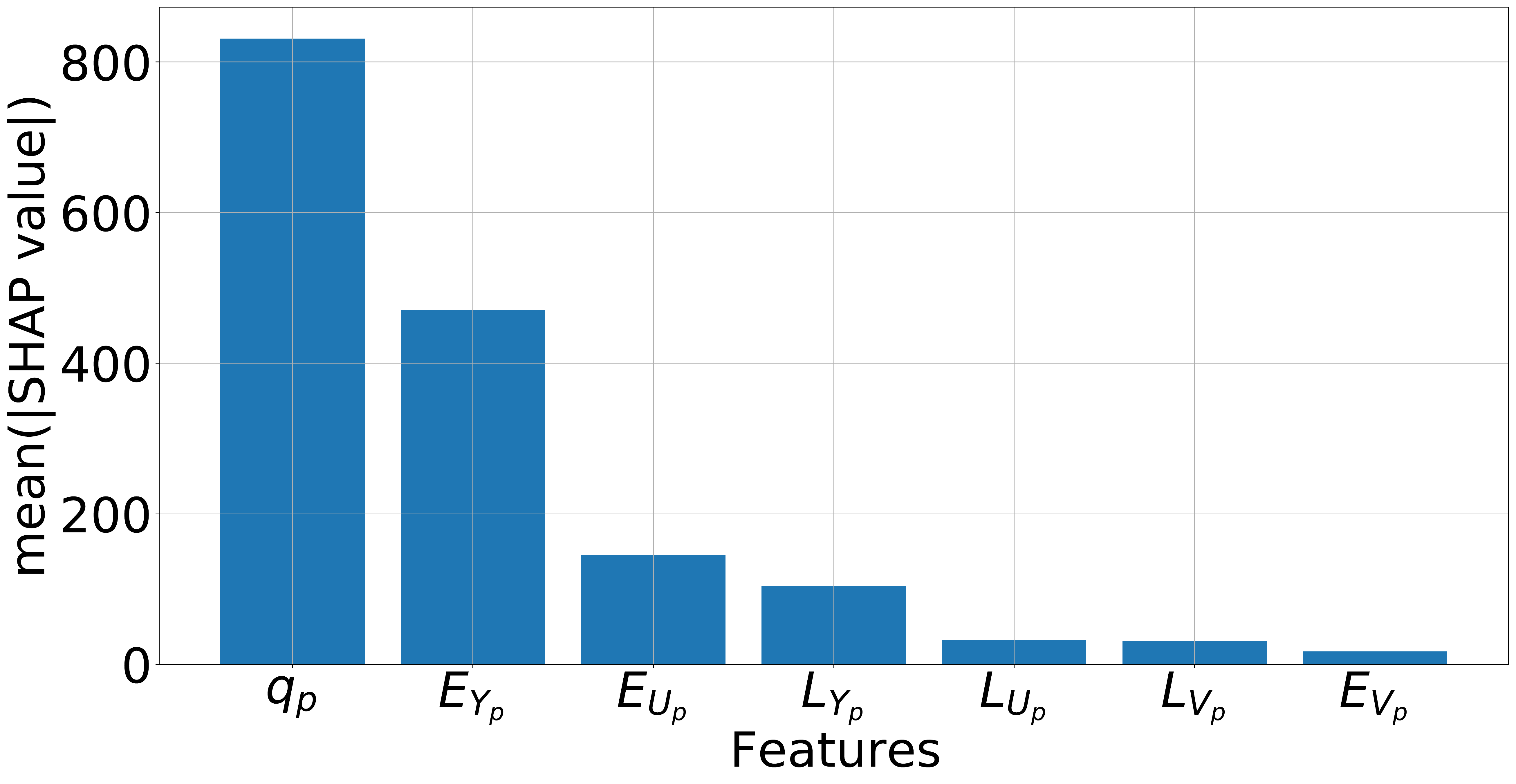}
\vspace{-0.5em}
\caption{Relative importance of features in bits prediction for I-frames.}
\label{fig:shap_i}
\end{figure}

\begin{figure}[t]
\centering
\begin{subfigure}{0.235\textwidth}
    \centering
    \includegraphics[width=\textwidth]{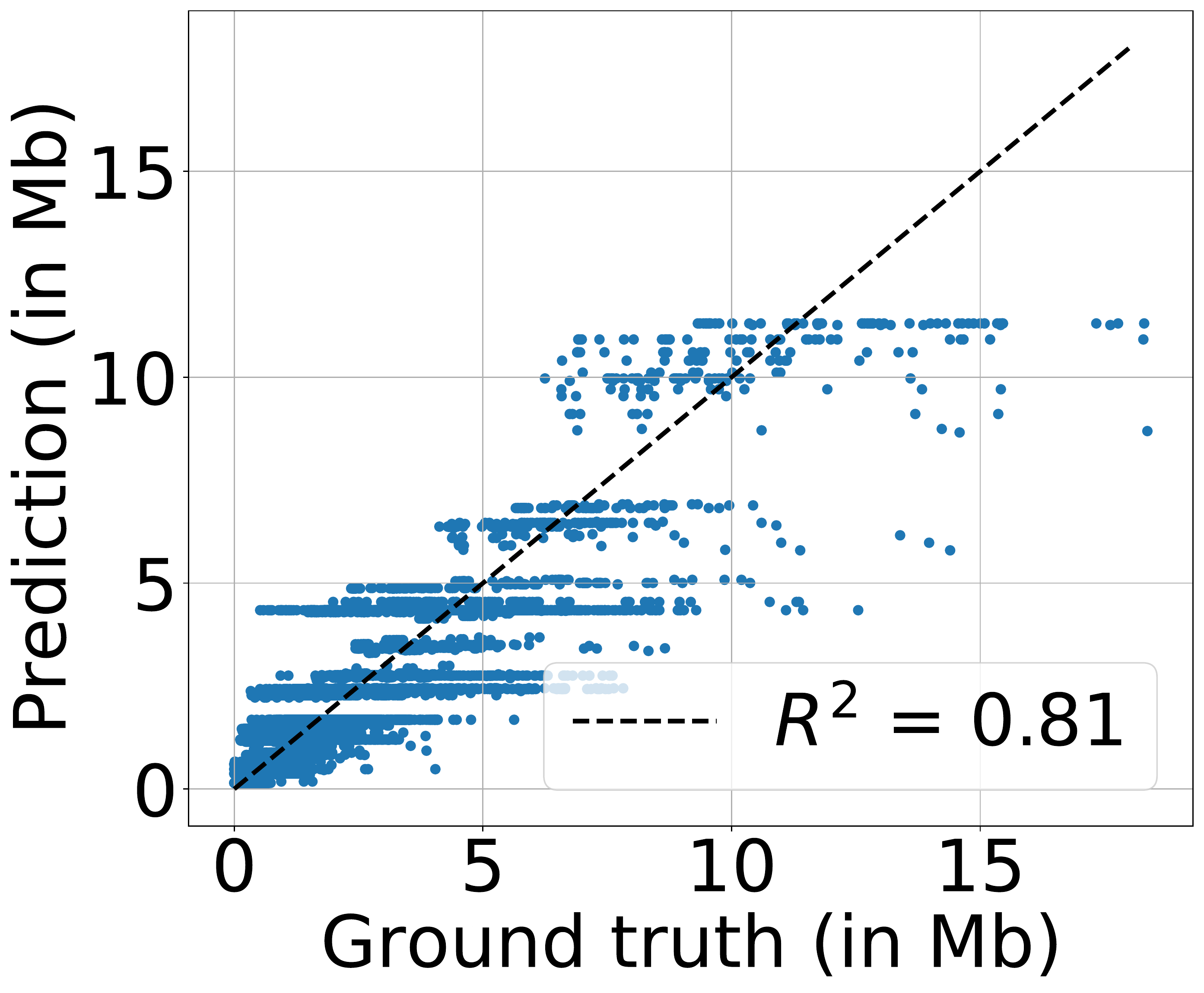}
    \caption{}
\end{subfigure}
\begin{subfigure}{0.235\textwidth}
    \centering
  \includegraphics[width=\textwidth]{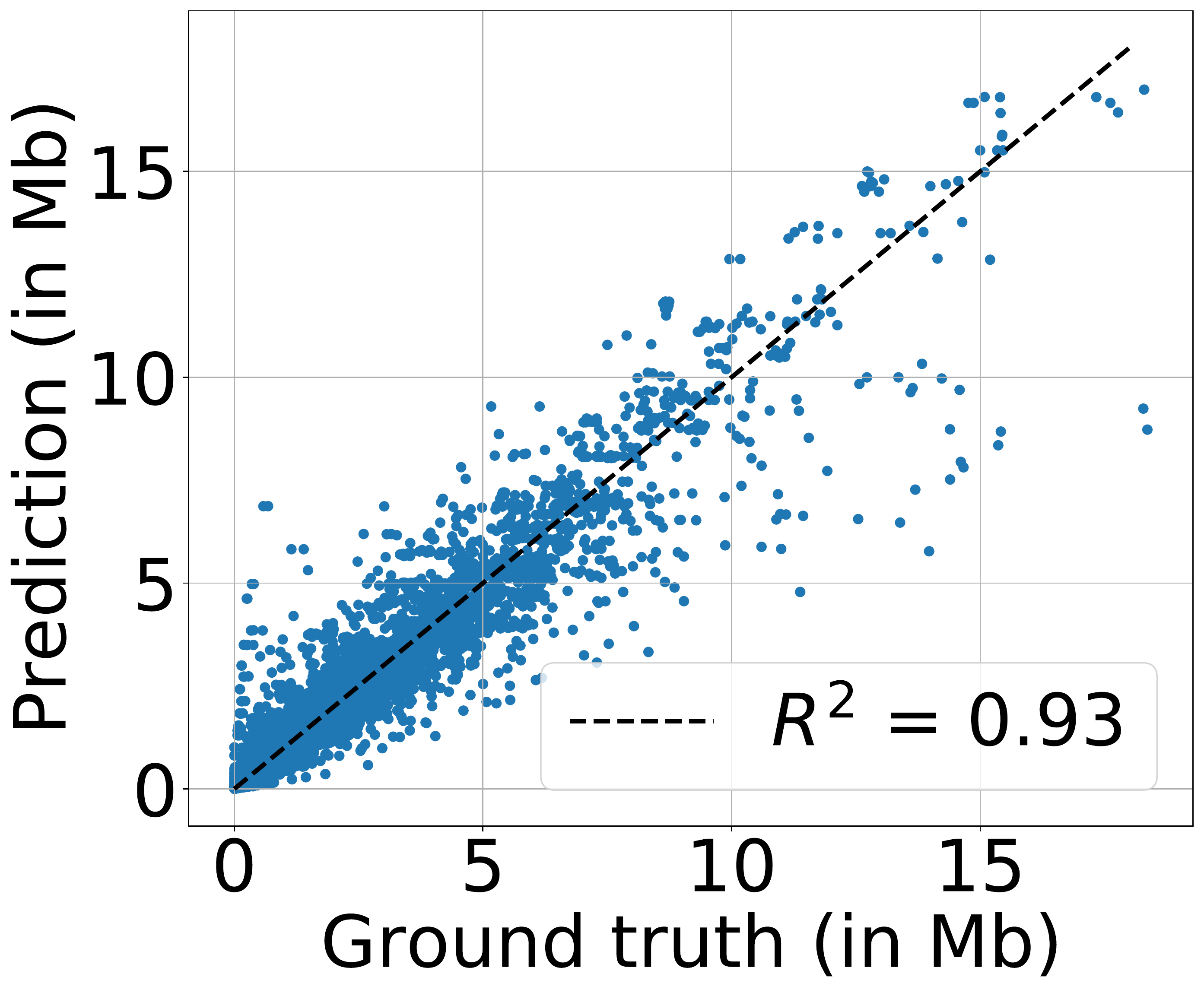}
    \caption{}
\end{subfigure}
\vspace{-0.75em}
\caption{Ground truth bits versus predicted bits per I-frame for \textit{max\_depth} values \textit{(a)} 4, and \textit{(b)} 12.}
\vspace{-0.8em}
\label{fig:scatter_plots_i}
\end{figure}

\paragraph*{Results:}
\label{sec:exp_results}
Fig.~\ref{fig:shap_i} shows the relative importance of the considered features in estimating bits of I-frames using SHAP values~\cite{shap_ref}. It is observed that $q_p$ is the most important feature, followed by the $E_{Y_p}$ feature. Table~\ref{tab:model_res_cons} analyzes the model size and the prediction accuracy for I-frames for various values of \textit{max\_depth}. When $max\_depth=12$, MAE is observed to be 172.59 kb, while $R^2$ and model size are 0.93 and 31.34 Mb, respectively. When $max\_depth=16$, MAE is reduced to 155.20 kb, while $R^2$ and model size are 0.93 and 139.87 Mb, respectively. Since further increasing $max\_depth$ does not improve the results, $max\_depth=12$ is used in the following experiments. The scatter-plot in Fig.~\ref{fig:scatter_plots_i} depicts the correlation between the ground truth and model predictions of $b$ for the considered values of $max\_depth$. A strong correlation between the predictions and ground truth is observed when $max\_depth$ is set as 12.

\begin{table}[t]
\caption{Bits estimator performance for various values of $max\_depth$.}
\centering
\resizebox{0.615\linewidth}{!}{
\begin{tabular}{l|c|c|c}
\specialrule{.12em}{.05em}{.05em}
\textit{max\_depth} & MAE & $R^2$ & Model Size \\
\specialrule{.12em}{.05em}{.05em}
4 & 409.31 kb & 0.81 & 0.22 Mb \\
8 & 217.25 kb & 0.92 & 3.21 Mb \\
12 & 172.59 kb & 0.93 & 31.34 Mb \\
16 & 155.20 kb & 0.93 & 138.87 Mb \\
\specialrule{.12em}{.05em}{.05em}
\end{tabular}}
\label{tab:model_res_cons}
\end{table}

\subsection{Rate Control Performance}

\label{ssec:integr}
\paragraph*{Experimental Setup:}

To evaluate the performance of the proposed method for the first pass of rate control~\cite{rc_ref}, the VCA and bits estimator components are executed offline, and the output of the bits estimator part is used for the second pass in VVenC. The described method is combined with VVenC~v1.7.0\footref{note:vvenc}~\cite{vvenc_ref}. The RD performance is evaluated using the all-intra configuration at \textit{faster} preset~\cite{preset_ref}, without temporal subsampling. A predefined target rate was obtained from CTC-like coding with a fixed QP for each test. The presented coding efficiency was measured in Bjøntegaard Delta (BD) rate differences using JVET’s CTC sequences for SDR classes A1 and A2~\cite{seq1_ref}, and Fraunhofer HHI’s public Berlin test set~\cite{seq2_ref}. Following the requirements of the proposed method, the test set is converted to 8-bit and 30 fps sequences. All evaluations are performed on an Intel Xeon E5-2697A v4 cluster with Linux OS and a GCC 7.3.1 compiler using eight CPU threads. The combined YUV BD-rate is calculated based on a weighted PSNR sum across all components~\cite{yuv_ref}:
\begin{equation}
\textit{PSNR}_\textit{YUV}= {\left(\frac{6~\textit{PSNR}_{Y}+\textit{PSNR}_{U}+\textit{PSNR}_{V}}{8}\right)}.
\label{eq:bdyuv}
\end{equation}
The anchor is VVenC encoding of the test sequences using a fixed-QP all-intra setting and \textit{faster} preset. The time required for the VCA and bits estimator steps was validated compared to the first pass of VVenC.  The experiment was performed single-threaded on the same simulation platform.

\begin{figure}[t]
    \centering
    \includegraphics[width=0.26\textwidth]{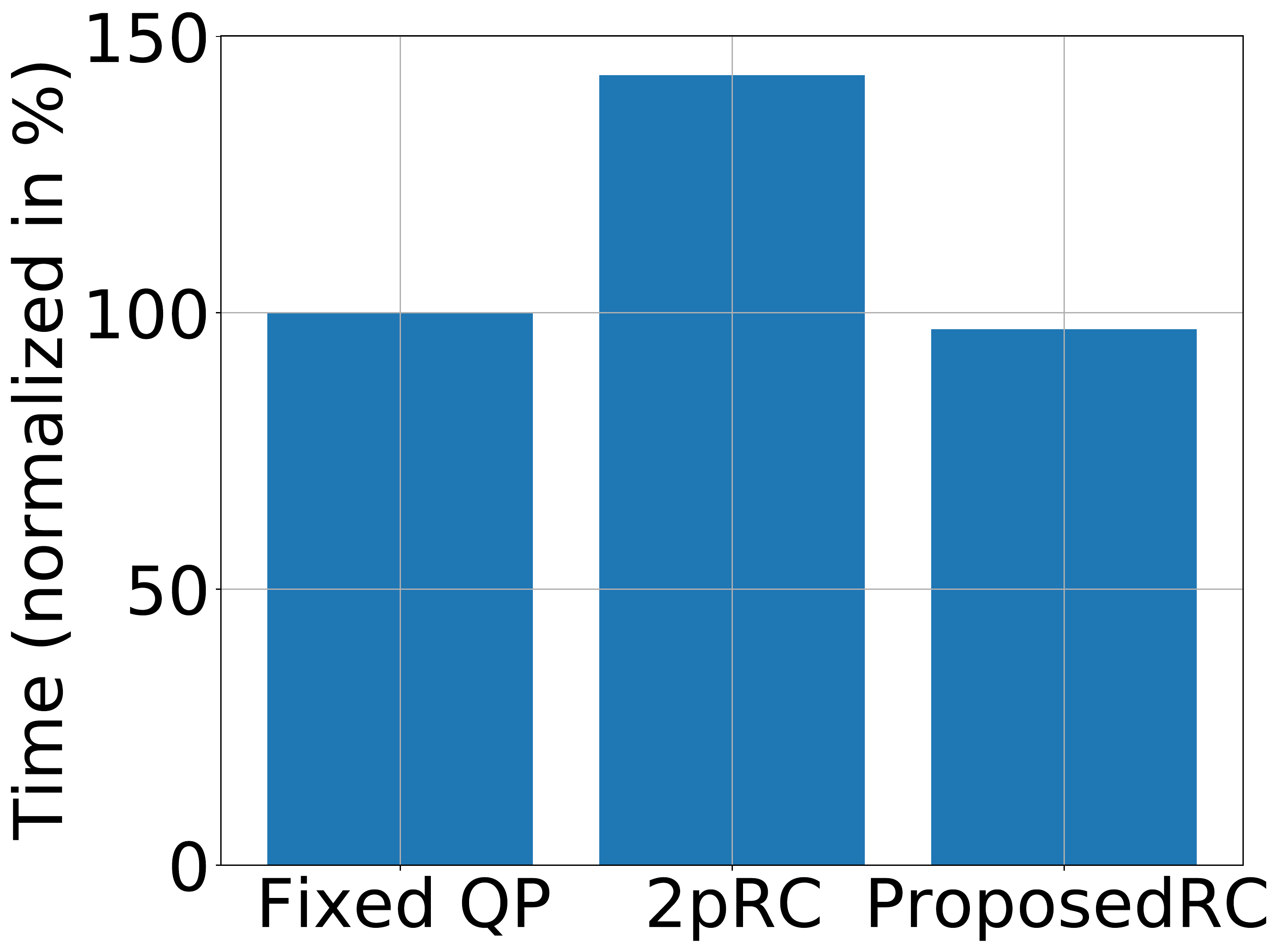}
\vspace{-0.3em}
\caption{Comparison of the overall encoding time using the considered rate control methods.}
\label{fig:2prc_time}
\end{figure}

\vspace{-0.4em}
\paragraph*{Results:}
In the first experiment, the runtime of the methods is compared. The analysis has shown that the first pass speed of 2pRC is 0.40 frames per second (fps), while the proposed method yields over 40 fps. Hence, the first pass of the proposed method is a hundred times faster than the first pass of 2pRC. The overall encoding time for the end-to-end application can be found in Fig~\ref{fig:2prc_time}. The encoding time of the proposed scheme is 32.17\% lower than 2pRC and 3\% lower than fixed QP encoding. 

The second experiment compares the rate-distortion (RD) performance of the methods. The results vary across the test sequences (\cf Table~\ref{tab:res_tests}); on average, the proposed scheme results in a 2\% BD-rate increase compared to VVenC's original 2pRC method. Hence, the proposed method achieves an efficiency close to the fixed QP reference and 2pRC of VVenC.

The third experiment analyzes the bitrate deviation. It is observed that the bitrate deviation is close to zero for all classes, indicating that the proposed method does not deteriorate the bitrate accuracy. The behavior is similar to the 2pRC in VVenC.

\begin{table}[t]
\caption{Results compared to the fixed-QP encoding using 2pRC and the proposed method (\cf Section~\ref{sec:proposed_alg}).} 
\centering
\resizebox{1.00\linewidth}{!}{
\footnotesize
\begin{tabular}{c|c|c|c|c}
\multirow{2}{*}{\textbf{Dataset}} & \multirow{2}{*}{\textbf{Sequence}} & \textbf{2pRC} & \textbf{Proposed RC} & \textbf{Noise} \\ 
 &  & $BD_{YUV}$[\%] & $BD_{YUV}$[\%] & $BD_{YUV}$[\%] \\ \hline
JVET UHD & Tango1	& 0.09  &	8.35  & 33.55\\
 & FoodMarket4	& 1.68  &	-0.64  & 17.98\\
 & Campfire	& 0.17  &	0.31  & 24.30\\
 & CatRobot &	0.18  &	0.81   & 22.94\\
 & DaylightRoad2	& 0.21  &	4.96  & 28.48\\
 & ParkRunning3 & 0.02  & -1.15  & 13.83\\
\hline
JVET UHD & Average & \textbf{0.39}  & \textbf{2.11}  & \textbf{23.51}  \\ 
\hline
Berlin set & BerlinCrossroadsCrop4K	& 0.17  &	0.44  & 27.57\\
 & ChestnutTreeCrop4K	& 0.26  &	0.34  & 18.69\\
 & March18thSquareCrop4K	& 0.11  &	7.25  & 15.62\\
 & NeptuneFountainCrop4K	& 0.93  &	2.67  & 19.09\\
 & OberbaumCrop4K	& 0.18  &	0.29  & 17.24\\
 & QuadrigaCrop4K	& 0.22  &	0.30 & 28.77\\
 & ReichstagIntoTreeCrop4K	& 0.38  & 2.57  & 18.04\\
 & SpreeCrop4K	& 1.23  &	2.15  & 23.40\\
\hline
Berlin set & Average & \textbf{0.44}  & \textbf{2.00}  & \textbf{21.05} \\ 
\hline
 & BerlinMix4K	& 0.06  & 2.69  & 14.38\\ 
 \hline
\end{tabular}}
\begin{flushright}
\footnotesize The sequences were resampled to 2160p 8bit 30fps.
\end{flushright}
\vspace{-1.75em}
\label{tab:res_tests}
\end{table}

Additionally, the performance of the proposed method was validated for significant changes in video content by testing a combined video sequence consisting of a concatenation of the eight Berlin sequences~\cite{seq2_ref}. The results in Table~\ref{tab:res_tests} for the BerlinMix4K sequence show an accuracy roughly comparable to the average of other sequences. 
The worst-case scenario is also reviewed to systematize the results and evaluate the operating points. To achieve the worst-case, gaussian white noise $\hat{b}_p= \mathcal{N}(\mu,\sigma^2)$, with mean $\mu$ and standard deviation $\sigma$ equal to the target per frame bits, simulated as the input to the second pass of the 2pRC encoding. The confrontation of rate control with the arbitrary noise signal allows us to validate the new approach and determine the limit of possible deterioration. The proposed rate control method achieves, on average, significantly lower compression losses ranging from slight gains of $-1.15\%$ to losses up to $8.35\%$, compared to noise with the degradation above $20\%$ on average and ranging up $33.55\%$. 
\vspace{-0.8em}
\section{Conclusions}
\label{sec:conclusion_future_dir}
\vspace{-0.7em}
A simplified first pass operation for all-intra 2pRC in VVenC, a practical \VVC encoder, has been presented. Around the preset \textit{faster}, the proposed method reduces the encoding time by 32.17\%, on average causing a 2\% BD-rate increase over the default 2pRC method, \ie requiring only 2\% more bits to produce the same objective quality. The proposed first pass is realized using the Video Complexity Analyzer (VCA) and an RF model to predict the required per-frame bits from the VCA features. This pre-analysis is used instead of a complete encoding with a reduced toolset used in VVenC. Especially for the preset \textit{faster}, the approach yields significant time savings of 32\%, achieving runtime on par with that of fixed QP encoding.

\balance
\setlength{\parskip}{0pt}
\setlength{\itemsep}{0pt}
\newpage
\bibliographystyle{IEEEtran}
\bibliography{references.bib}
\balance
\end{document}